\documentstyle[aas2pp4, amsmath,xspace, enumerate]{article}

\newcommand{\etal}{{\it et al. \@\xspace}}
\slugcomment{Submitted to Astrophysical Journal Letters}
\lefthead{Nilsson et al.}
\righthead{Isotropic CMB in isotropic universes}

\begin{document}
\title{An almost isotropic cosmic microwave temperature does not
  imply an almost isotropic universe}

\author{U. S. Nilsson}
\affil{Department of Applied Mathematics \\ University of Waterloo
  \\Waterloo, Ontario N2L 3G1, Canada}
\author{C. Uggla}
\affil{Department of Engineering
  Sciences, Physics and Mathematics\\
  University of Karlstad\\ S-651 88
  Karlstad, Sweden}
\author{J. Wainwright}
\affil{Department of Applied Mathematics \\ University of Waterloo
  \\Waterloo, Ontario N2L 3G1, Canada}
\and\author{W. C. Lim}
\affil{Department of Applied Mathematics \\ University of Waterloo
  \\Waterloo, Ontario N2L 3G1, Canada}

\begin{abstract}
In this letter we will show that, contrary to what is widely believed,
an almost isotropic cosmic 
microwave background (CMB) temperature does not
imply that the universe is ``close to a Friedmann-Lemaitre
universe''. There are two important manifestations of anisotropy in
the geometry of the universe, (i) the anisotropy in the overall
expansion, and (ii) the intrinsic 
anisotropy of the gravitational field, described by the Weyl curvature
tensor, although the former usually receives more attention than the
latter in the astrophysical literature. Here we consider a class of
spatially homogeneous models for which the anisotropy of the CMB
temperature is within the 
current observational limits but whose Weyl curvature is not negligible,
i.e. these models are not close to isotropy even though the CMB
temperature is almost isotropic.

\keywords{Cosmology: theory -- cosmic microwave background}
\end{abstract}
\twocolumn
Observations indicate that the temperature of the CMB is isotropic to
a remarkable degree. This observation is widely believed to indicate
that the universe is almost isotropic in the sense that it is ``close to
a Friedmann-Lemaitre (FL) model'', at least since the time of last
scattering. Support for this belief is provided by a fundamental
result of Ehlers, Geren \& Sachs (\cite{art:Ehlersetal1968}), known as
the EGS theorem, which states
that if the CMB temperature were exactly isotropic about every point
in 
spacetime, the universe would have to be exactly an FL model. This
result is of course not directly applicable to the real universe
since the CMB temperature is not exactly isotropic. It is thus of
considerable 
interest to note that Stoeger \etal (\cite{art:Stoegeretal1995}) were
able, by making what might seem to be reasonable assumptions, to
extend the 
results of Ehlers \etal by replacing the word ``exactly'' by
``almost'', thus obtaining an ``almost'' EGS theorem. More precisely,
they showed, given certain hypotheses, 
that if the CMB temperature is measured 
to be almost isotropic in a spacetime region of an expanding universe,
then the universe is close to an FL model {\it in that region}.

Before proceeding further, it is necessary to give a precise meaning
to the phrase ``close to an 
FL model''. We follow the approach given in Wainwright \& Ellis
(\cite{book:WainwrightEllis1997}, section 2.4), but focus
primarily on the question of anisotropy rather than inhomogeneity.
To quantify the deviation of the universe from an FL model as
regards to anisotropy, the shear
tensor $\sigma_{ab}$ (see Wainwright \& Ellis
\cite{book:WainwrightEllis1997}, page 18 for the definition) of  
the cosmological fluid and the Weyl curvature  
tensor $C_{abcd}$ are of primary importance since both are zero in an
FL model. The shear tensor describes anisotropy in the overall
expansion of the universe, while the Weyl curvature tensor represents
the  
``free gravitational field'', describing for example, tidal forces. It
is related to time derivatives of the shear tensor (see Wainwright \&
Ellis 
\cite{book:WainwrightEllis1997}, Eqs.\ (1.43) and (1.47)). Requiring
the above quantities to be small in an
anisotropic ever--expanding cosmology is not sufficient since these
quantities are not dimensionless. Moreover, the shear, for example,
tends to zero at late times 
irrespective of whether the model isotropizes or not. The 
appropriate quantities are the dimensionless ratios formed by
normalizing with the Hubble scalar $H$
(see Wainwright \& Ellis \cite{book:WainwrightEllis1997}, page 18 for
the definition), thus measuring the dynamical 
importance of the different variables with respect to the overall
expansion of the universe. This insight dates back to the important paper
on observations
in cosmology by Kristian \& Sachs (\cite{art:KristianSachs1966}, see
page 398), in which they showed that various geometrical quantities, 
including the Weyl curvature tensor, can in
principle be restricted by observations of distant galaxies (see their
Table 1 for a list of their so-called anisotropy and inhomogeneity
parameters). We therefore introduce the parameters $\Sigma$ and ${\cal
  W}$ according to,
\begin{equation}
  \label{eq:Sigmadef}
  \Sigma^2 =
  \frac{\sigma_{ab}\sigma^{ab}}{6H^2}\ , \quad
  {\cal W}^2 = 
  \frac{E_{ab}E^{ab} + H_{ab}H^{ab}}{6H^4}\ ,
\end{equation}
where $\sigma_{ab}$ is the shear tensor and $E_{ab}$ and $H_{ab}$ are
the electric and magnetic parts of the Weyl curvature tensor
respectively. They are defined by
\begin{equation}
 E_{ac} = C_{abcd}u^b u^d\ ,\quad 
 H_{ac} = \tfrac{1}{2}\epsilon_{ab}\!^{ef}C_{efcd}u^b u^d\ .
\end{equation}
where $u^a$ is the 4-velocity of the cosmological fluid. We refer to 
$\Sigma$ as the {\it shear parameter} and to ${\cal W}$ as the {\it
  Weyl 
parameter}. For non-tilted spatially homogeneous perfect fluid
models, a zero shear tensor implies that the Weyl curvature tensor is
zero, and thus characterizes the FL models. However, restricting the
Hubble-normalized shear to be small does {\it not} guarantee that the
Hubble-normalized Weyl curvature is small since derivatives of
$\sigma_{ab}$ need not be small compared to $H^2$. Therefore a 
necessary condition for
the universe to be close to an FL model is 
that the Hubble-normalized shear {\it and} the Hubble-normalized Weyl
curvature are small. In terms of the previously introduced
anisotropy parameters, this necessary condition becomes
\begin{equation}
  \Sigma << 1\ , \ {\cal W} << 1\ .
\end{equation}

To characterize the anisotropies of the CMB temperature, a
spherical harmonic expansion of the temperature of the CMB sky,
\begin{equation}
  \frac{\Delta T}{T} = \frac{T\left(\theta, \varphi\right) -
\bar{T}}{\bar{T}} = \sum
  a_{lm}Y_{lm}(\theta, \varphi)\ , 
\end{equation}
is often used. In the above expression, $\bar{T}$ is the mean temperature
of the CMB sky and
$\left(\theta,\varphi\right)$ are the usual spherical angles. The
multipoles, 
\begin{equation}
  a_l = \left(\sum_{m=-l}^{l}a_{lm}^2\right)^{1/2}\ ,
\end{equation}
describe, coordinate independently, the deviation of the CMB
temperature from perfect isotropy. The dipole anisotropy, $a_1$, is
commonly interpreted as describing 
the motion of the solar system with respect to the rest frame of the 
CMB, see e.g. Lineweaver \etal (\cite{art:Lineweaveretal1996}),
making the quadrupole, $a_2$, the lowest multipole to describe ``true''
anisotropies. The observational bounds on the quadrupole 
and octupole are of the order (Stoeger \etal \cite{art:gebbie})
\begin{equation}
  \label{eq:gebbie}
  a_2\approx10^{-5} \ , \quad a_3\approx10^{-5}\ .
\end{equation}

The line of research initiated by Stoeger \etal 
(\cite{art:Stoegeretal1995}) into the implications of CMB temperature
observations has been extended by Maartens \etal in a series of papers
(Maartens \etal \cite{art:Maartensetal1995a}, \cite{art:Maartensetal1995b},
  \cite{art:Maartensetal1996}) to give explicit 
bounds on various anisotropy and inhomogeneity parameters. It turns
out that to first order, only the first three 
multipoles: the dipole, quadrupole and octupole, play a direct role,
via the Einstein-Boltzmann equations, in limiting the anisotropy and
inhomogeneity parameters. 
Specifically, Maartens \etal (\cite{art:Maartensetal1995b}) deduced
general inequalities from which we infer the following 
upper bounds on the present values of the shear and Weyl parameters, 
using the observational bounds on $a_2$ and $a_3$ in Eq.\
(\ref{eq:gebbie}):
\begin{equation}
  \label{eq:bounds}
  \Sigma_{\rm o}<10^{-4}, \quad {\cal  W}_{\rm o}  < 10^{-3}\ .
\end{equation}
These follow from Eqs. (31), (35), and (36) in Maartens \etal
(\cite{art:Maartensetal1995b}) using our Eq.\ (\ref{eq:Sigmadef}). It
is important to note, however, that these 
upper bounds are valid only if a number of hypotheses about the
behavior of the multipole moments in the universe since last
scattering are true. The essential hypotheses can be stated as
follows: 

\begin{enumerate}[$H_1:$]
\item the Copernican principle holds approximately,
\item the multipole moments have never been larger than they are
    today,
\item the dimensionless time and spatial derivatives of the
  multipoles are bounded by the multipoles themselves (Maartens \etal
  \cite{art:Maartensetal1995b}, assumptions C1' and C2').
\end{enumerate}

\noindent
The observation of the CMB temperature multipoles at our present
position and at 
the present time, in conjunction with hypotheses $H_1$ and $H_2$,
gives bounds on the multipoles in a spacetime region between last
scattering and the present, encompassing all fundamental observers
with whom we have been in causal contact with since last
scattering (Maartens \etal \cite{art:Maartensetal1995a}, Eqs. (42)
and (44)). The derivatives of the multipoles are not directly 
accessible to us (Maartens \etal \cite{art:Maartensetal1996}, page L9)
since to obtain this information would require 
observations over time and space intervals on cosmological scales.
A close examination of the paper by Maartens \etal
(\cite{art:Maartensetal1995b}, Eqs.\ (24) - (29)) shows that all of
the upper bounds on the various anisotropy and inhomogeneity
parameters depend on hypothesis $H_3$ and are thus not, in practice, 
observationally testable. Thus, {\it if hypothesis $H_3$ is not
  satisfied 
then the CMB temperature observations} , in conjunction with the
analysis of Maartens \etal (\cite{art:Maartensetal1995b}), {\it do not
  impose upper bounds on 
$\Sigma$ and ${\cal W}$}, and hence do not establish that the universe
is close to 
FL. It is clearly of interest to ask whether one can construct model 
universes which illustrate this possible limitation of the CMB
temperature observations. 

In this letter we exhibit such a class of cosmological models, namely
the 
spatially homogeneous non-tilted Bianchi type VII$_0$ dust models 
in which the CMB is treated as a test field.\footnote{An explicit
  introduction of a radiation fluid not interacting with the dust will
  not change the results.} To obtain the present CMB temperature
pattern, the photon energies are integrated 
along the null geodesics that connect points of emission on the
surface of last scattering with the event of observation at the
present time. It will be assumed that the process of 
matter-radiation decoupling in the early universe took place
instantaneously and that the CMB photons travel to us along null
geodesics from a redshift of $z\approx 1100$. If $T_{\rm e}$ is the
temperature of the CMB at the surface of last scattering, the observed
temperature $T_{\rm o}$ is given by (Nilsson \etal 
\cite{art:Nilssonetal1998})  
\begin{equation}\label{eq:temp2}
  T_{\rm o} = T_{\rm e}\exp \left\{ -\int_{\tau_{\rm e}}^{\tau_{\rm o}} 
   \left[1 + \Sigma_{ab}(\tau)K^a(\tau)
         K^b(\tau)\right]d\tau\right\}\ , 
 \end{equation}
where the subscripts o and e refer to observation and emission
respectively. Here $K^a$ are the directional cosines of a particular 
photon path and $\Sigma_{ab}=\sigma_{ab}/H$ is the Hubble--normalized 
shear tensor. The parameter $\tau$ is a
dimensionless time variable defined in terms of the Hubble scalar by
\begin{equation}
  \label{eq:taudef}
        \frac{dt}{d\tau} = H^{-1}\ ,
\end{equation}
where $t$ is the cosmological clock time.  In an ever--expanding model,
$\tau$ will assume all real values with $\tau\rightarrow +\infty$ at  
late times. The $\tau$-dependence of $\Sigma_{ab}$ and $K^a$ is found
by integrating the field equations and 
the null geodesic equations, which for spatially homogeneous models
form a coupled system of ordinary differential equations (Nilsson 
\etal \cite{art:Nilssonetal1998}). If the universe is {\it assumed} to be 
close to an FL model, then 
$\Sigma_{ab}$ is small and Eq.\ (\ref{eq:temp2}) can be linearized in
$\Sigma_{ab}$. Also, to lowest order, the directional cosines of the
null geodesics in the corresponding background FL model can be
used. For spatially homogeneous models, this approach has been used by
various authors to obtain bounds on the 
Hubble normalized shear from CMB temperature observations. We refer,
for example 
to Collins \& Hawking (\cite{art:CollinsHawking1973a}), 
Barrow (\cite{art:Barrow1985}), Bunn \etal (\cite{art:Bunnetal1996})
and Kogut \etal (\cite{art:Kogutetal1997}). Since we do {\it not} assume
a priori that the models we are considering are close
to an FL model, the full field 
equations and the exact null geodesic equations were integrated
numerically to obtain the CMB temperature. 

To quantify the anisotropy of the Bianchi type VII$_0$ dust models and
hence their deviation from an FL model, we use the shear parameter
$\Sigma$ and the Weyl parameter ${\cal W}$ introduced previously.
It has been shown by Collins \& Hawking
(\cite{art:CollinsHawking1973b}) that Bianchi VII$_0$ dust models
initially close to the flat FL model isotropize at late times in the
sense that $\Sigma \rightarrow 0$ as $\tau\rightarrow
+\infty$. A more detailed analysis by Wainwright \etal
(\cite{art:Wainwrightetal1998}) showed that for any non-tilted
Bianchi VII$_0$ dust model, the anisotropy parameters satisfy
\begin{equation}
 \lim_{\tau\rightarrow +\infty} \Sigma =0\ ,\quad
 \lim_{\tau\rightarrow +\infty}{\cal W} = {\cal W}_\infty\ ,
\end{equation}
at late times. Here ${\cal W}_\infty$ is a constant whose value can be
any positive 
number depending on the initial conditions. In other words, {\it
  although 
the models isotropize as regards the shear, the Weyl curvature
remains dynamically significant}. This fact can be understood
heuristically by considering the asymptotic form of the shear
parameter $\Sigma$ ,
\begin{equation}
  \Sigma \approx C{\rm e}^{-\tfrac{\tau}{2}}\left|
  \cos\left(\tfrac{4}{H_0l_0}{\rm e}^{\tfrac{\tau}{2}}  +
  \psi_0\right)\right|\ ,\quad \tau \rightarrow +\infty\ ,
\end{equation}
where $C, l_0, H_0$ and $\psi_0$ are constants. This asymptotic form
follows 
from  Eq.\ (3.29), Eq.\ (3.33) and Eq.\ (C.10) in Wainwright \etal
(\cite{art:Wainwrightetal1998}). The decaying 
exponential ensures that $\Sigma$ is small but the fact that
$\Sigma$ is highly oscillatory at late times generates a dynamically
significant Weyl curvature.

We have performed a variety of
numerical simulations to calculate anisotropy patterns of the CMB
temperature in Bianchi VII$_0$ models. Initial conditions are imposed at 
the present time $\tau_{\rm o}$ and then the field equations
and null geodesics equations are integrated backwards to the time 
of last
scattering $\tau_{\rm e}$. Then Eq.\ (\ref{eq:temp2}) is used to obtain
the ratio $T_{\rm o}/T_{\rm e}$. The 
interval $\tau_{\rm o} - \tau_{\rm e}$, corresponding to a redshift of
$z\approx 1100$,  is approximately 7. For a typical simulation, with
present day values of the anisotropy parameters being $\Sigma_{\rm o}
= 3\cdot 10^{-5}$ and ${\cal W}_{\rm o} = 2$,  we obtain a quadrupole
of the order $a_2 \approx 7 \cdot 
10^{-6}$ and an octupole of the order $a_3 \approx
7\cdot10^{-8}$. This model thus 
satisfies the observational constraint Eq.\ (\ref{eq:gebbie}) but
violates the theoretical upper bounds of Eq.\ (\ref{eq:bounds}) as
regards the Weyl parameter. Hence, in the case of the present models,
one or several 
of the hypotheses $H_{1}, H_2, H_3$ are not valid. Since the models we
are considering are spatially homogeneous, 
hypothesis $H_1$ is valid and the calculations show that hypothesis
$H_2$ is valid as well. It is thus, as expected, hypothesis $H_3$ that
fails. 

It is of interest to note that the octupole is considerably smaller
than the quadrupole for these models. This is not a
coincidence. Analytic considerations suggest that for Bianchi VII$_0$
models, a {\it pure} quadrupole pattern is obtained asymptotically at
late times. Doroshkevich 
\etal (\cite{art:Doro}) have also given heuristic arguments that
support this conclusion. It would be useful if this result could be
proved and 
understood physically in a simple way since it is perhaps surprising.
In order to investigate if it is
possible to obtain detailed observationally compatible CMB
temperature patterns using these models, one will have to consider
density perturbations 
of these models. Will they give rise to similar CMB temperature maps
as those obtained from density perturbations of FL models? Does a 
dynamically important Weyl curvature give rise to a different large
scale structure and if so, does this in turn affect the CMB
temperature?  

In this letter we have shown that there exists cosmological models,
specifically spatially homogeneous non-tilted Bianchi type VII$_0$
dust models, that are not close to any FL model even though the
temperature of the CMB is almost isotropic in the sense that the
observational bounds on the quadrupole and octupole are satisfied.
It is also clear why this
does not contradict the results of Maartens \etal
(\cite{art:Maartensetal1995b}) since one of the hypotheses
on which their analysis is based is not valid for these models. 
Thus, the existence of these models suggests that hypothesis $H_3$
may not always be appropriate. It should be pointed out that the
present class of models may very well be ruled out by observations
other than 
those of the CMB temperature. However, this is exactly the
point. There is indeed a need for other cosmological observations.

\acknowledgments
We thank Roy Maartens for commenting on an earlier draft of the
manuscript and Tim Gebbie for helpful discussions. This research was
supported in part by a grant from the 
Natural Sciences \& Engineering Research Council of Canada (JW),
the Swedish Natural Research Council (CU), G{\aa}l\"ostiftelsen (USN),
Svenska Institutet (USN), Stiftelsen Blanceflor (USN) and the
University of Waterloo (USN, WCL).

\end{document}